\documentstyle[preprint,tighten,eqsecnum,aps,floats,psfig,epsfig,graphics]{revtex}

\newcommand{\fd}{fluc\-tu\-a\-tion-dis\-si\-pa\-tion }
\newcommand{\beq}{\begin{equation}}
\newcommand{\beqa}{\begin{eqnarray}}
\newcommand{\eeq}{\end{equation}}
\newcommand{\eeqa}{\end{eqnarray}}

\newcommand{\dpar}{\partial}
\renewcommand{\d}{{\rm d}}
\newcommand{\en}{\varepsilon}

\begin{document}

\title{
Aging at Criticality in Model C Dynamics
}
\author{P.~Calabrese and A.~Gambassi}
\address{Scuola Normale Superiore and INFN,
Piazza dei Cavalieri 7, I-56126 Pisa, Italy. 
\\
{\bf e-mail: \rm
{\tt calabres@df.unipi.it},
{\tt andrea.gambassi@sns.it}
}
}

\date{\today}

\maketitle

\begin{abstract}
We study the off-equilibrium two-point critical response and 
correlation functions for the relaxational dynamics  with a coupling to a 
conserved density (Model C) of the $O(N)$ vector model. 
They are determined in an $\epsilon=4-d$ expansion for vanishing momentum.
We briefly discuss their scaling behaviors and the associated 
scaling forms are determined up to first order in $\epsilon$. 
The corresponding \fd ratio has a non trivial 
large time limit in the aging regime and, up to one-loop order, it is
 the same as that of the Model A for the physically relevant 
case $N=1$.
The comparison with predictions of local scale invariance is also discussed.
\end{abstract}



\section{Introduction}
\label{intr}

Non-equilibrium dynamics of statistical systems is currently under intensive 
theoretical investigation, and new dynamical behaviors have been recently 
discovered in models of disordered systems. One of the most striking of 
them is  {\it aging}, i.~e. a persistence of the system in a 
non-equilibrium state even after a macroscopic time has elapsed since
the latest perturbation acting on it. As a consequence, there 
is no ``memory loss'' of the thermal history of the system and its 
response to an external field, for example, will depend on it. 
This fact is commonly observed in glassy systems~\cite{glass,review}. 
It has been pointed out~\cite{ckp-94}, however, that this kind of 
behavior may be also observed in critical non--disordered models. 
In these cases the presence of slow-relaxing modes could keep the system 
in a non-equilibrium state even asymptotically for large times.  
Consider, indeed, a system in a generic configuration and, at time $t=0$, 
bring it in contact with a thermal bath at a given 
temperature $T$. The resulting relaxation process is 
characterized by a transient behavior with off-equilibrium evolution, 
for $t<\tau_R$, and a stationary equilibrium evolution for $t>\tau_R$, 
where $\tau_R$ is the relaxation time. In the former the behavior of the 
system is expected to depend on initial conditions, while in the 
latter time homogeneity and time reversal symmetry (at least in the absence 
of external fields) are recovered and such a dependence is lost; 
fluctuations are thus described in terms of ``equilibrium'' dynamics.

In the following we focus on ferromagnetic systems
$\varphi$, quenched at their critical 
temperature $T_c$ \cite{foot} for $t=0$~(interesting behaviors are observed 
also in the case of non-instantaneous quench, i. e. for time dependent thermal 
bath \cite{phr-02}). 
A convenient way of describing dynamics is to study 
two-time response and correlation functions. The former is usually defined as 
$R_{\bf x} (t,s)=\delta\langle \varphi_{\bf x} (t)\rangle/ \delta h (s)$, 
where $\varphi$ is the magnetic order parameter, $h$ is a small external 
field 
applied at time $s > 0$ in the point ${\bf x}={\bf 0}$, and 
$\langle \cdot \rangle$ stands for the mean over the stochastic dynamics. 
The latter, instead,  is defined as the order parameter correlation function 
$C_{\bf x}(t,s)=\langle \varphi_{\bf x}(t)\varphi_{\bf 0}(s)\rangle$.

If the system does not reach the equilibrium, the response and correlation 
functions will 
depend both on $s$~(the ``age'' of the system, also called ``waiting time'') 
and on the observation time $t$. 
To characterize the distance from equilibrium of an aging system, 
evolving at a fixed temperature $T$, the 
\fd ratio~(FDR) is usually introduced \cite{ck-94,ckp-94}:
\beq
X_{\bf x}(t,s)=\frac{T\, R_{\bf x}(t,s)}{\dpar_s C_{\bf x}(t,s)} \; .
\label{dx}
\eeq
When $t$ and the waiting time $s$ are both greater than $\tau_R$, 
the dynamics is homogeneous in time and time-reversal invariant so that 
the \fd theorem can be applied, leading to $X_{\bf x}(t,s)=1$.
This is no longer true in the aging regime \cite{ckp-94}.
It has been argued that the long time limit of the FDR at criticality
\beq
X^\infty=\lim_{s\to\infty}\lim_{t\to\infty}X_{{\bf x}=0}(t,s) \ ,
\label{xinfdef}
\eeq
is a novel {\it universal} quantity of non-equilibrium critical 
dynamics \cite{gl-000,gl-00,gl-02b}.
Correlation and response functions have been exactly computed for a Random 
Walk, a free Gaussian field, and a two-dimensional XY model at 
zero temperature and the value $X^\infty=1/2$ 
has been found \cite{ckp-94}. In the case of the
$d$-dimensional Spherical Model \cite{gl-00}, 
one dimensional Ising-Glauber chain \cite{lz-00,gl-000}
and two- and three- dimensional Ising Model, investigated by Monte Carlo 
simulations~\cite{gl-00}, $X^\infty$ has values ranging 
between $0$ and $1\over 2$.

Field-theoretical methods have been proven a powerful tool for 
the computation of {\it universal} quantities (such as critical exponents) 
in critical phenomena (for an updated review see Ref.~\cite{pv-02}).
In this framework the problem of critical relaxation from a 
macroscopically prepared initial state has been analyzed since some 
years, and a new universal exponent associated with it has been introduced 
as a consequence of an additional time-surface renormalization~\cite{jss-89}. 

We would take advantage of these previous works to compute the critical FDR
and the associated {\it universal} scaling functions for mesoscopic models of 
dynamics, overcoming most of the analytical difficulties encountered in 
the exact solutions of models with aging dynamics.
In Refs.~\cite{cg-02a} and~\cite{cg-02b} this problem has been addressed for 
the dissipative dynamics (Model A of Ref.~\cite{HH}) of the $O(N)$ 
ferromagnetic model, whereas the purely dissipative dynamics of the diluted 
Ising model has been analyzed In Ref. \cite{cg-02c}. 
Here we consider the $O(N)$ model dynamically coupled 
to a conserved density (Model C of Ref.~\cite{HH}).
Physical realizations of this models are, e.g., intermetallic 
alloys \cite{gfl-97}, adsorbed layers on solid substrates \cite{bkl-82}
and supercooled liquid \cite{t-99}.
Also the deterministic microcanonical $\varphi^4$ model \cite{micr,zst-99}
is believed to be in the Model C universality class since the order 
parameter is coupled to the conserved energy \cite{kc-02}.

The paper is organized as follows.
In Section~\ref{secC} Model C is introduced and 
its scaling forms are discussed.
In Section \ref{seconeloop} we derive the first order contribution in an 
$\epsilon$-expansion to the response and correlation functions
for all values of $s$ and $t$ and we derive the FDR up to the same order. 
Finally in Section~\ref{disc} we discuss our results stressing their 
relevance for the issue (of applicability) of local scale invariance.

\section{Model C}
\label{secC}

Let us consider the relaxational dynamics of an $N$-component 
field $\varphi({\bf x},t)$ coupled to a noncritical conserved 
density $\en({\bf x},t)$. This system may be described by means of the 
following coupled stochastic Langevin equations~(Model C of Ref.~\cite{HH})
\beqa
\label{lang}
\dpar_t \varphi ({\bf x},t)&=&-\Omega 
\frac{\delta \cal{H}[\varphi,\en]}{\delta \varphi({\bf x},t)}+\xi({\bf x},t) \; , \\
\dpar_t \en({\bf x},t)&=& \Omega\rho\nabla^2 
\frac{\delta \cal{H}[\varphi,\en]}{\delta \en({\bf x},t)}+\zeta({\bf x},t) \; ,
\eeqa
where $\cal{H}[\varphi,\en]$ is the Landau-Ginzburg Hamiltonian for the 
fields $\varphi$ and $\en$\ with a coupling term between them
\beq
{\cal H}[\varphi,\en] = \int \d^d x \left[
\frac{1}{2} (\nabla \varphi )^2 + \frac{1}{2} r_0 \varphi^2
+\frac{1}{4!} g_0 \varphi^4 +  \frac{1}{2} \en^2 +  \frac{1}{2}\gamma_0 \en\varphi^2 \right] ,\label{lgw}
\eeq
where $\Omega$ and $\rho$ are the kinetic coefficients, $r_0\propto T-T_c$,
$g_0$ and $\gamma_0$ the bare coupling constants, 
$\xi({\bf x},t)$ and $\zeta({\bf x},t)$  zero-mean stochastic 
Gaussian noises with
\beqa
\langle \xi_i({\bf x},t) \xi_j({\bf x}',t')\rangle&=& 2 \Omega \, \delta({\bf x}-{\bf x}') \delta (t-t')\delta_{ij} ,\\ 
\langle \zeta({\bf x},t) \zeta({\bf x}',t')\rangle &=& - 2 \rho \,\Omega\, \nabla^2 \delta({\bf x}-{\bf x}') \delta (t-t')\; .
\eeqa  
The coupling between $\en({\bf x},t)$ and $\varphi({\bf x},t)$ does not change 
the static properties of the latter as it can be seen by computing the 
effective 
Hamiltonian for the $\varphi$ field (see Ref.~\cite{ZJ-book}). 
Moreover $\en$-field static correlation functions are related to 
$\varphi^2$-field correlation functions. 

Dynamical correlation functions, generated by the Langevin 
equations~(\ref{lang}) and averaged over the noises $\xi$ and $\zeta$, 
may be obtained by means of the field-theoretical action \cite{bjw-76,ZJ-book} 
\beq
S[\varphi,\tilde{\varphi},\en,\tilde{\en}]= \int \d t \int \d^dx 
\left[ \tilde{\varphi} \partial_t\varphi +
\Omega \tilde{\varphi} \frac{\delta \mathcal{H}[\varphi,\en]}{\delta \varphi}-
\tilde{\varphi} \Omega \tilde{\varphi}  
+\tilde{\en} \partial_t\en -
\rho\,\Omega \tilde{\en} \nabla^2\frac{\delta \mathcal{H}[\varphi,\en]}{\delta \en} +
\tilde{\en} \rho\,\Omega \,\nabla^2\tilde{\en} \right],\label{mrsh}
\eeq
where $\tilde{\varphi}({\bf x},t)$ and  $\tilde{\en}({\bf x},t)$ are the 
response fields associated with $\varphi({\bf x},t)$ and 
$\en({\bf x},t)$, respectively. It is easy to read from Eq.~(\ref{mrsh}) 
and~(\ref{lgw}) the interaction vertices, given by 
$-\Omega g_0 \tilde{\varphi} \varphi^3/3!$, as in the case of Model~A, 
$ - \Omega \gamma \en \tilde{\varphi}\varphi$ and $\rho \Omega \gamma \,\varphi^2\nabla^2\tilde{\en}/2$.

In Ref.~\cite{jss-89,oj-93} this formalism was extended to deal with 
relaxation of the system from a  macroscopically prepared initial state. 
To take into account the effect of such initial condition on the dynamics 
described by Eq.~(\ref{mrsh}) one has also to average over the possible 
initial configurations of both the order parameter 
$\varphi_0({\bf x})=\varphi({\bf x},t=0)$ and the conserved density 
$\en_0({\bf x})=\en({\bf x},t=0)$
with a probability distribution  $e^{-H_0[\varphi_0,\en_0]}$ given 
by~\cite{oj-93}
\beq
H_0[\varphi_0]=\int\! \d^d x\, \left[ \frac{\tau_0}{2}(\varphi_0({\bf x})-u({\bf x}))^2 + \frac{1}{2 c_0}(\en_0({\bf x})- v({\bf x}))^2\right].
\eeq
This specifies an initial state $u({\bf x})$ for $\varphi({\bf x},t)$ and 
$v({\bf x})$ for $\en ({\bf x},t)$ with correlations proportional to $\tau_0^{-1}$ and $c_0$, 
respectively. 
Response and correlation functions may
be obtained, following standard methods \cite{bjw-76,ZJ-book}, by a 
perturbative expansion of the functional weight
$e^{-(S[\varphi,\tilde{\varphi},\en,\tilde\en]+H_0[\varphi_0,\en_0])}$. 
An initial condition with long-range correlations may lead to a different 
universality class, as e.g. shown for the $d$-dimensional spherical model with 
non-conservative dynamics \cite{ph-02}.

The propagators~(Gaussian two point correlation and response functions) 
of the resulting theory are \cite{oj-93} 
\beqa
\langle \tilde{\varphi_i}({\bf q},s) \varphi_j(-{\bf q},t) \rangle_0 =& 
\delta_{ij} R^0_q(t,s)=&\delta_{ij} \,\theta(t-s) G(t-s),\label{Rgaux}\\
\langle \varphi_i({\bf q},s) \varphi_j(-{\bf q},t) \rangle_0 =&
\delta_{ij} C^0_q(t,s)=& {\delta_{ij} \over q^2+r_0}\left[ G(|t-s|)+\left(\frac{r_0 +q^2}{\tau_0}-1
\right) G(t+s)\right], \label{Cgaux}
\eeqa
where
\beq
G(t)=\displaystyle{e^{-\Omega (q^2+r_0) t}} \label{GG},
\eeq
and \cite{oj-93}
\beqa
\langle \tilde{\en}({\bf q},s) \en(-{\bf q},t) \rangle_0 =& 
R^0_{\en,q}(t,s)=& \theta(t-s) G_\en(t-s),\label{Rgauxe}\\
\langle \en({\bf q},s) \en(-{\bf q},t) \rangle_0 =&
 C^0_{\en,q}(t,s)=& G_\en(|t-s|)+ (c_0 - 1) G_\en(t+s), \label{Cgauxe}
\eeqa
with
\beq
G_\en(t)=\displaystyle{e^{-\rho\Omega (q^2+r_0) t}} \label{GGe}.
\eeq
As in the case of Model~A and Model~B, it has been shown that $\tau_0^{-1}$ 
is irrelevant (in the renormalization group sense) so that
 we set $\tau_0^{-1} = 0$\cite{jss-89,oj-93}.

\subsection{Scaling forms}
\label{secscalingC}

When a ferromagnetic system is quenched from a disordered initial state to
its critical point, the correlation length grows as $t^{1/z}$, where $z$ is
the dynamical critical exponents \cite{HH} and $t$ the time elapsed since the
quench. So in  momentum space, applying standard scaling arguments, the 
universal two time $(s,t)$ response and correlation functions depend 
only on the two products $q^z \, t$ and $q^z\, s$, where $q$ is the external
momentum.

In particular general renormalization group argument suggest the scaling forms 
\cite{jss-89,oj-93}
\beqa
\Omega R_{\bf q=0}(t,s)&=&A_R (t-s)^a (t/s)^\theta F_R(s/t) \, ,
\label{Rscalform}\\
C_{\bf q=0}(t,s)&=&A_C s (t-s)^a (t/s)^\theta F_C (s/t)\, ,\label{Cscalform} 
\eeqa
where $R_{\bf q}(t,s)$ and $C_{\bf q}(t,s)$ are the Fourier transforms~(with
respect to ${\bf x}$) of $R_{\bf x}(t,s)$ and $C_{\bf x}(t,s)$ respectively,
$a=(2-\eta-z)/z$ \cite{foot1.2}, and $\theta$ is the initial-slip exponent of 
response function\cite{jss-89,oj-93}. 
The functions $F_C(v)$ and $F_R(v)$ are universal provided
one fixes the nonuniversal normalization constant $A_R$ and $A_C$
to have $F_i(0)=1$. 

In~\cite{cg-02a} the following  quantity, related to the FDR, 
was introduced  in momentum space
\beq
{\cal X}_{\bf q}(t,s)=
{\Omega R_{\bf q}(t,s)\over \dpar_s C_{\bf q}(t,s)}.\label{Xq}
\eeq
It has been argued that the zero-momentum limit 
\beq
{\cal X}_{\bf q=0}^{\infty}=
\lim_{s\to\infty}\lim_{t\to\infty} {\cal X}_{\bf q=0}(t,s)\, ,
\label{Xq2}
\eeq
is equal to the same limit of the FDR~(\ref{xinfdef}) for ${\bf x=0}$, i.e. 
${\cal X}_{\bf q=0}^\infty=X^\infty$ to all orders~\cite{cg-02a}. 
This fact allows an easier perturbative computation (in momentum space) 
of the new universal quantity $X^\infty$. 
Combining scaling forms and previous definitions, we find
\beq
{\cal X}_{{\bf q}=0}^\infty(t,s)=\lim_{s\to\infty}\lim_{t\to\infty}
{T_c R_{{\bf x}=0}(t,s)\over \dpar_s C_{{\bf x}=0}(t,s)}=
{A_R\over A_C (1-\theta)}\, .
\label{xinffromsf}
\eeq

In recent works the notion of local scale invariance has been introduced
as an extension of anisotropic or dynamical scaling (see Ref.~\cite{henkel-02}
and references therein). Assuming the covariance of the response function 
under a suitable subgroup of the constructed group of local scale 
transformations, it has been argued that \cite{henkel-02}
\beq
R_{\bf x}(t,s)=R_{\bf x=0}(t,s) \Phi (|{\bf x}|/(t-s)^{1/z}),
\label{henkelpred}
\eeq
where \cite{hpgl-01}
\beq
R_{\bf x=0}(t,s)= {\cal A}_R  (t-s)^{a'} (t/s)^\theta ,
\eeq
and $\Phi(u)$ is a function whose explicit and convergent series expansion
is known \cite{henkel-02}.
Fourier transforming Eq. (\ref{henkelpred}) and setting ${\bf q}=0$
one could obtain the strong prediction $F_R(s/t)=1$.
For the correlation function and its derivative no analogous result exists.

\section{One-loop FDR}
\label{seconeloop}

In this Section we compute the non-equilibrium critical two-point 
response and correlation functions for the Model~C up to one-loop order,
for vanishing external momentum. 
We use here the method of renormalized field theory in the minimal 
subtraction scheme. 
The breaking of time homogeneity gives rise to some technical problems in
the renormalization procedure in terms of one-particle irreducible correlation
functions \cite{jss-89} so our computation is done in terms of connected 
functions.

\begin{figure}
\epsfig{width = 15cm, file=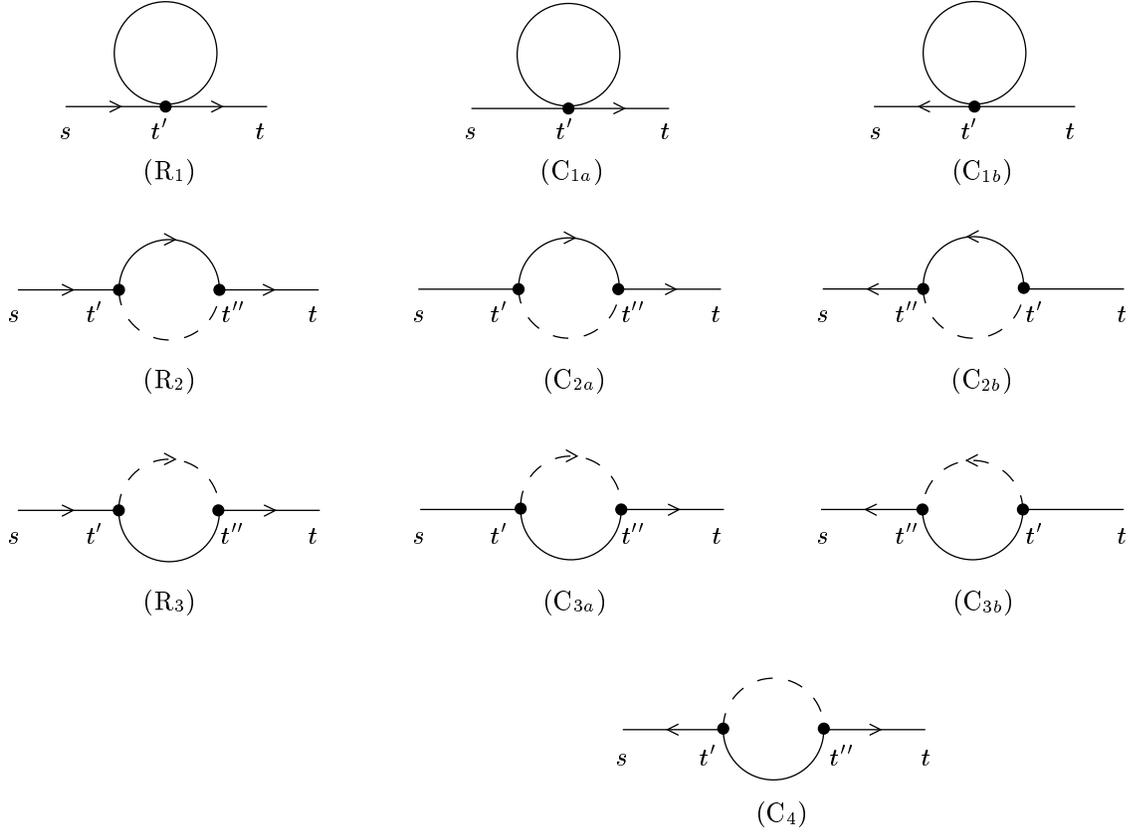} 
\caption{Feynman diagrams contributing to the one-loop order-parameter 
response~((R$_1$), (R$_2$), (R$_3$)) and 
correlation function~((C$_{1a,b}$), (C$_{2a,b}$), (C$_{3a,b}$), (C$_{4}$)). 
Response functions are drawn as lines with arrows going from the early time
to the later one, whereas
correlators bear no arrow lines. Solid (dotted) lines refer to the
order parameter $\varphi$ (to the conserved density $\en$).
}
\label{fig}
\end{figure}

At one-loop order we have to evaluate, taking also into account causality 
\cite{bjw-76}, the ten Feynman diagrams depicted
in Figure~\ref{fig}, three for the response function 
((R$_1$), (R$_2$) and (R$_3$)) and seven for
the correlation one ((C$_{1a,b}$), (C$_{2a,b}$), (C$_{3a,b}$), and (C$_{4}$)). 

In terms of them we have
\beqa
R_q(t,s) =  R_q^0(t,s)&-& {N+2 \over 6} g_0 ({\rm R}_1) + \Omega^2 \gamma^2 ({\rm R}_2) + \rho \Omega^2 \gamma^2 ({\rm R}_3) +O(g_0^2, g_0\gamma^2,\gamma^4)\;,\nonumber\\
C_q(t,s)= C_q^0(t,s)&-& {N+2 \over 6} g_0 [ ({\rm C}_{1a})+({\rm C}_{1b})]+ 
\Omega^2 \gamma^2 [({\rm C}_{2a})+({\rm C}_{2b})] \\
&+& \rho \Omega^2 \gamma^2 [({\rm C}_{3a})+({\rm C}_{3b})] +
\rho \Omega^2\gamma^2 ({\rm C}_{4})
+ O(g_0^2,g_0\gamma^2,\gamma^4)\;.\label{int}
\eeqa

In order to evaluate the FDR at criticality we set, in
this perturbative expansion, $r_0=0$~(massless theory).
We also set $\tau_0^{-1}=0$, since it is an irrelevant variable \cite{jss-89}, 
and $\Omega=1$ to lighten the notations.
The first step in the calculation of the
diagrams is the evaluation of the critical ``bubbles'' $B_{1c}(t)$, 
$B_{2c}(t',t'')$, $B_{3c}(t',t'')$, and $B_{4c}(t',t'')$, 
i.e. the one-particle irreducible parts common to diagrams depicted on 
the first, second, third and fourth line of Figure~\ref{fig}, respectively.
We have, in generic dimension $d$ \cite{cg-02a}
\beq
B_{1c}(t)= \int \! {\d^dq \over (2 \pi)^d} \; C^0_q(t,t)=-
{1\over d/2-1}{(2 t)^{1-d/2} \over (4 \pi)^{d/2}}=- N_d {\Gamma(d/2-1)\over 2^{d/2}} t^{1-d/2},
\label{bolla1}
\eeq
where $N_d= {2/ (4 \pi)^{d/2} \Gamma(d/2)}$.
Given we are interested in the value of the FDR~(\ref{Xq}) for 
\mbox{${\bf q} = {\bf 0}$} we evaluate, in the following, all diagrams for 
vanishing external momentum. Then
for $B_{2c}(t,s)$, $B_{3c}(t,s)$  and $B_{4c}(t,s)$ we have
\beqa
B_{2c}(t,s)&=& \int \! {\d^dq \over (2 \pi)^d} \;  R^0_q(t,s)\, C^0_{\en,q}(t,s) = \nonumber \\
& =& \theta(t-s) [4\pi\Omega(1+\rho)]^{-d/2} \left[ (t-s)^{-d/2} + (c_0 -1)(t - \kappa\, s)^{-d/2}\right] \; , \label{bolla2}\\
B_{3c}(t,s)&=& \int \! {\d^dq \over (2 \pi)^d} \; q^2 \,R^0_{\en,q}(t,s)\, C^0_q(t,s) = \nonumber\\
& =& \theta(t-s) [4\pi\Omega(1+\rho)]^{-d/2} \left[ (t-s)^{-d/2} - (t+\kappa \, s)^{-d/2}\right] \; , \label{bolla3} \\
B_{4c}(t>s,s) &=& \int \! {\d^dq \over (2 \pi)^d} \; C^0_{\en,q}(t,s)\, C^0_q(t,s) = \frac{N_d}{2}\Gamma(d/2-1) [\Omega(1+\rho)]^{1-d/2} \cdot \nonumber\\& \cdot& \left\{ (t-s)^{1-d/2}- (t+\kappa\, s)^{1-d/2} + (c_0-1)[(t- \kappa\, s)^{1-d/2} - (t + s)^{1-d/2}\right\} \; , \label{bolla4}
\eeqa
where $\kappa = (1-\rho)/(1+\rho) < 1$ 
(given that, for Model C to make sense, $\rho > 0$). 
Expression~(\ref{bolla4}) for $B_{4c}(t,s)$\ is valid only for $t>s$, 
that for $s>t$ is easily found, given the symmetry property 
$B_{4c}(t,s) = B_{4c}(s,t)$. Once critical bubbles have been determined, it 
is easy to compute each diagram in Figure~\ref{fig}.

Performing the required integrations and 
expanding in powers of $\epsilon$ we find, for the bare response function,
\beqa
R_{q=0}(t,s)= - \frac{2 \tilde{\gamma_0}^2}{1 + \rho} \frac{1}{\epsilon} + 1+ \left[\tilde{g}_0{N+2\over 24} - \tilde{\gamma_0}^2\frac{1 + \rho^2 - c_0}{2 \rho (1 + \rho)} \right] \ln {t\over s} +\nonumber\\
-\frac{\tilde{\gamma_0}^2}{1 + \rho} \ln[\Omega (t-s)]
-\tilde{\gamma_0}^2 c_0 \frac{1}{1 - \rho^2} \ln\frac{1 - \kappa v}{1 - \kappa} +\tilde{\gamma_0}^2 {\mathcal R}(s/t;\rho) +\nonumber\\
+ O(\epsilon^2,\tilde{g}_0^2, \epsilon\tilde{g}_0, \tilde{\gamma}^4, \tilde{\gamma}^2\tilde{g}_0,\epsilon\tilde{\gamma}^2) 
\; ,\label{Reps}
\eeqa
where
\beq
 {\mathcal R}(v;\rho) =   
 -\frac{\rho}{1 - \rho^2} \ln\frac{1 + \kappa v}{2} +
   \frac{1}{1 - \rho^2} \ln\frac{1 - \kappa v}{2\rho} 
 -\frac{1}{1 + \rho}
\; ,
\eeq
and for the correlation function
\beqa
C_{q=0}(t,s)= -\frac{4 \tilde{\gamma_0}^2 \Omega s}{1 + \rho} \frac{1}{\epsilon} +2 \Omega s
\left\{ 1+ 
\tilde{g}_0{N+2\over 12}+ \left[\tilde{g}_0{N+2\over 24} - \tilde{\gamma_0}^2 \frac{1 + \rho^2 - c_0}{2 \rho(1+\rho)}\right]\ln\frac{t}{s} +\right.\nonumber\\
\left.- \frac{\tilde{\gamma_0}^2}{1 + \rho} \ln[\Omega t]+ \tilde{\gamma_0}^2 \left[
- c_0 {\mathcal C}_1(\frac{s}{t};\rho) -c_0 {\mathcal C}_2(\rho) 
- \frac{\ln[1 + \rho]}{1 + \rho} + {\mathcal C}_2(\rho) + \frac{1}{\rho} {\mathcal C}_2(\frac{1}{\rho})\right.\right. \nonumber \\
\left.\left. +
 {\mathcal C}_1(\frac{s}{t};\rho) - {\mathcal C}_1(-\frac{s}{t};\rho) \right]\right\} + O(\epsilon^2,\tilde{g}_0^2, \epsilon\tilde{g}_0, \tilde{\gamma}^4, \tilde{\gamma}^2\tilde{g}_0,\epsilon\tilde{\gamma}^2)\;,\label{Ceps} 
\eeqa
where we assumed $t > s$ and we introduced $\tilde{g_0}=N_d g_0$, $\tilde{\gamma_0}=N_d \gamma_0$ and the functions
\beqa
{\mathcal C}_1(v;\rho) &=& \frac{1+v}{2v(1+\rho)} \ln[1 + v] -\frac{1 - \kappa v}{2 v (1+\rho)\kappa^2} \ln[1 - \kappa v] \; ,
\\
{\mathcal C}_2(\rho) &=&  - \frac{\ln[1 - \kappa]}{(1 - \rho)^2} - \frac{1}{(1 - \rho) \rho}  \; .
\eeqa
The first one is defined for $-1<v<1/\kappa$ and $\rho\neq 1$ 
(we are interested only in the case $\rho\ge 0$). 
We note that contributions to Eq. (\ref{Ceps}) coming from ${\mathcal C}_i$
are regular in the limit $\rho \rightarrow 1$. 

The previous expressions for $R_{q=0}$ and $C_{q=0}$ have 
simple poles in $\epsilon$, so renormalization of the bare parameters 
is required. We use the minimal subtraction scheme in order to render 
renormalized quantities finite for $\epsilon \rightarrow 0$. At one-loop 
order it is sufficient to perform the following 
renormalizations~\cite{bd-75,ZJ-book}
\beq
\left\{\begin{array}{lcl}
\tilde\varphi &\mapsto& \tilde Z^{-1/2} \tilde\varphi\\
\Omega &\mapsto& \tilde Z^{-1/2} \Omega 
\end{array}
\right. \  ,\ \mbox{with} \ \ \tilde Z = 1 - \frac{4\tilde{\gamma}^2}{1+\rho} \frac{1}{\epsilon} + O(\tilde{\gamma}^4,\tilde{\gamma}^2\tilde{g}, \tilde{g}^2) \ \ , 
\eeq
to render two-point functions finite, since $Z=1+O(\tilde{g}^2)$ as known from
statics \cite{ZJ-book}. 

Let us briefly recall the scenario of fixed points for out-of-equilibrium
Model C ~\cite{bd-75,ZJ-book,oj-93}.
The fixed-point values for the couplings $g$ and $\gamma$  are determined 
only by the statics.
We have $\tilde{g}^* = \tilde{g}^*_A + 6\tilde{\gamma}^{2*}$, where 
$\tilde{g}^*_A=6\epsilon/(N+8) + O(\epsilon^2)$\ is the fixed-point value
of the coupling constant for Model A \cite{ZJ-book}.

The value of $\gamma$ at the infrared stable fixed point depends on the sign
of the specific-heat exponent $\alpha$:
\beq
\tilde{\gamma}^{2*}=
\left\{\begin{array}{cll}
0\; , & \mbox{stable for } \alpha <0 \, ,& \mbox{ case (I)}\, ,\\
\frac{4-N}{N(N+8)}\epsilon + O(\epsilon^2) \;, &  \mbox{stable for }
\alpha >0 \, , & \mbox{ case (II)}\, ,
\end{array}\right.
\eeq
in the case (I), the dynamics of the conserved density decouples from that of 
the order parameter and we get back to Model A (at least asymptotically). 
At the leading order in $\epsilon$-expansion we have, 
for the $O(N)$ model \cite{pv-02},
\beq
\alpha = \frac{4-N}{2(N+8)}\epsilon + O(\epsilon^2) \; ,
\eeq   
thus the truly Model C dynamical fixed point is stable for $N<4+O(\epsilon)$.
In three dimensions, numerical calculations shows \cite{pv-02}
that $\alpha$ is negative already for $N=2$, so that the Model C dynamics
may be realized only for the three-dimensional Ising model ($N=1$) that has
positive $\alpha$ \cite{pv-02} (the two-dimensional Ising model has $\alpha=0$
and the values of dynamical critical exponents for Model C are still 
debated \cite{2d}).

As far as $\rho$\ is concerned we have three possible stable fixed points 
determined by equilibrium dynamics\cite{bd-75}
\begin{itemize}
\item[(a)] $\rho^* = \infty$, stable for 
$N>N_1(\epsilon)=4-[15/4+3/2 \log (4/3)] \epsilon+O(\epsilon^2)$;
\item[(b)] $\rho^* = 2/N -1 + O(\epsilon)$, stable for  
$N < 2+ C \epsilon |\ln (\epsilon)|$ and for $N_2<N<N_1$, where 
$N_2(\epsilon)=4-[7/2+3\ln(4/3)]\epsilon+O(\epsilon^2)$;
\item[(c)] $\rho^* = 0$, which governs the 
critical behavior in the complement of the two regions,
but it is a peculiar limit~\cite{ZJ-book}.
\end{itemize}
Finally, regarding the out-of-equilibrium dynamics,
it has been shown that, whenever $\alpha>0$, the fixed point value for $c$\ 
is $c^* = 0$~\cite{oj-93}.

We focus our attention on the only relevant stable fixed point of the model, 
i.e. (IIb), for which
\beq
\tilde{g}^*= {24\over N(N+8)} \epsilon +O(\epsilon^2).
\eeq

Taking into account scaling forms~(\ref{Rscalform}) and~(\ref{Cscalform}), 
we find the well-known critical exponents for 
Model C~\cite{ZJ-book,oj-93}~(some of 
these results have been corrected at two-loop order in Ref.~\cite{fm-02}) 
\beq
\left\{
\begin{array}{cll}
\theta &= {\displaystyle \tilde{g}^*{N+2\over 24} - \tilde{\gamma}^{2*} \frac{1 + \rho^{*2}}{2 \rho^*(1+\rho^*)}} + O(\epsilon^2)\; 
=& {N^2-8N+10\over 2(N-2)(N+8)} \epsilon+ O(\epsilon^2)\, ,\\
{\displaystyle \frac{2-\eta-z}{z}}&= 
- {\displaystyle \frac{\tilde{\gamma}^{2*}}{1 + \rho^*}}  + O(\epsilon^2)\; =
& -{4-N\over 2(N+8)}\epsilon+ O(\epsilon^2)\, ,
\end{array}
\right.
\eeq
and the scaling functions $F_R$ and $F_C$ are easily identified in 
Eqs.~(\ref{Reps}) and~(\ref{Ceps}) with $c^* = 0$
\beqa
F_R(v) &=&
1 + \tilde{\gamma}^{*2} [{\mathcal R}(v;\rho^*)-{\mathcal R}(0;\rho^*)] 
+  O(\epsilon^2)\; ,\\
F_C(v) &=& 1+ \tilde{\gamma}^{2*} \Bigg[\frac{1}{1 + \rho^*} \ln(1-v) + 
{\mathcal C}_1(v;\rho^*) - {\mathcal C}_1(-v;\rho^*) \Bigg]+ O(\epsilon^2) \; .
\eeqa
In particular substituting fixed point values, we obtain the scaling form
(we remember that $0\leq v \leq 1$ and $N < 2+O(\epsilon)$ so that no worries 
about the sign in the argument of the second logarithm arise)
\beq
F_R(v)=1+{4-N\over 4(N+8)(N-1)}\epsilon
\left[(N-2)\ln (1+(N-1)v)+N \ln (1-(N-1)v)\right]+ O(\epsilon^2)\; ,
\eeq
and, for the physically relevant case of $N=1$,
\beq
F_R(v)=1-\epsilon {v\over 6}+ O(\epsilon^2)\; ,
\eeq
that displays a correction to the mean-field value already at one-loop order
(at variance with Model A~\cite{cg-02a}).

We are now in the position to evaluate the FDR for Model C.
We first note that its Gaussian expression is the same as that of
Model~A as far as $\varphi$ and $\tilde\varphi$ are concerned and of Model~B 
(with some straightforward changes due to 
noncritical behavior of the conserved field) for $\en$ and  $\tilde\en$.
In order to evaluate the $\varphi$-field FDR  we 
compute the derivative with respect to $s$ of the two-time correlation 
function and consider its ratio  with the response one:
\beqa
\frac{1}{2}{\cal X}^{-1}_{q=0}(t,s)=
1+\tilde{g}^* \frac{N+2}{24} + \tilde{\gamma}^{2*} \left[
\frac{1}{1+\rho^*}\ln \frac{1-s/t}{1 + \rho^*} + {\mathcal C}_2(\rho^*) + \frac{1}{\rho^*} {\mathcal C}_2(\frac{1}{\rho^*}) 
+ \frac{1+\rho^{*2}}{2\rho^*(1+\rho^*)} +\right. \nonumber \\ \left.
+ {\mathcal C}_1(\frac{s}{t};\rho^*) - {\mathcal C}_1(-\frac{s}{t};\rho^*)
- {\mathcal R}(\frac{s}{t},\rho^*)
+\frac{s}{t}\left[ \partial_1{\mathcal C}_1(\frac{s}{t},\rho^*) + \partial_1{\mathcal C}_1(-\frac{s}{t},\rho^*)\right]
\right]+ O(\epsilon^2)\; ,
\label{X-1} 
\eeqa
where $\dpar_1$ stands for the derivative with respect to the first argument.
Note that, at variance with the one-loop FDR of Model A, this result depends 
on $s/t$.

In the limit $t\rightarrow\infty$, $s$ fixed, we find an $s$-independent 
expression,
\beqa
\frac{1}{2}\left({\cal X}^{\infty}_{q=0}\right)^{-1}=
1+\tilde{g}^* \frac{N+2}{24} + \tilde{\gamma}^{2*} \left[
\frac{2\rho^*}{(1+\rho^*)(1-\rho^*)^2}\ln\frac{(1+\rho^*)^2}{4\rho^*}- \frac{1+\rho^*}{2\rho^*}\right]
+ O(\epsilon^2)\; .
\label{X-1inf} 
\eeqa
Taking into account the fixed point values of couplings we find
\beq
 {\cal X}^\infty_{q=0} = \frac{1}{2}\left\{1 +\frac{4-N}{N(N+8)}\epsilon\left[\frac{N(N-1)}{(4-N)(2-N)} + \frac{N^2(2-N)}{4(N-1)^2}\ln[N(2-N)]\right] \right\} \ + O(\epsilon^2)\; .
\eeq
For $N=1$, which is the physically relevant case into which Model C is 
non trivial, the result is exactly the same as in Model A, 
${\cal X}^\infty_{q=0} = 1/2(1-\epsilon/12)+O(\epsilon^2)$, i.e. the presence 
of a coupled conserved density does not affect the value of 
${\cal X}^\infty_{q=0}$, at least up to one-loop order.  

In the $\epsilon$-expansion for 
$N>N_1(\epsilon)=4-[15/4+3/2 \log (4/3)] \epsilon+O(\epsilon^2)$ a fixed point 
with $\rho^*=\infty$ governs the critical behavior of the systems, but it
probably disappears in three dimensions. In this limit we find~(considering 
always $\alpha>0$ to ensure $c_0=0$)
\beq
\frac{1}{2}\left({\cal X}^\infty_{q=0}\right)^{-1} = 
1+\tilde{g}^*_A\frac{N+2}{24}+ {N\over4}\tilde{\gamma}^{2*} +O(2\mbox{-loop}) \; .
\eeq
Once again, as it happens in all the models that have been considered so far 
in the literature, the loop 
corrections lead to an FDR that is less than the mean field value $1/2$.

\section{Conclusions}
\label{disc}

In this work we considered the off-equilibrium properties of the $N$-vector
model coupled to a conserved energy density (Model C) in the framework of 
the field theoretical $\epsilon$-expansion.
We computed up to the first order in $\epsilon$ the critical FDR
as a function of the waiting time $s$ and of the observation
time $t$.
In the long-time limit, for the physically relevant case of one component
(Ising model) this ratio has the same value as in purely dissipative Model A.
Higher-loop calculations may clarify whether this is only a coincidence at
one-loop or it is a deeper property.

We also obtain the $O(\epsilon)$ expression for the response and 
correlation function for vanishing external momentum. In both the cases we
found corrections to the mean-field forms. Thus the result for the response 
function apparently disagrees with the prediction of local scale 
invariance (see Ref.~\cite{henkel-02} for an exhaustive introduction), i.e. 
$F_R(v)=1$. The same disagreement was 
already noted at two-loop order in the response function of 
Model A \cite{cg-02b}.
In that case, however, the presence of a very small prefactor in the correction
makes very hard the detection of this effect both in experiments and in 
Monte Carlo simulations. In  the present case, instead, for $N=1$, the 
correction in
\beq
F_R(v)=1-\epsilon {v\over 6}+ O(\epsilon^2)\; ,\label{is}
\eeq
should be large enough to be detectable. A Monte Carlo simulation of Model C 
could be helpful to clarify the nature of this disagreement. 
In Ref. \cite{cg-02b} we stressed that some problems in the comparison between
Eq. (\ref{is}) and the predictions of local scale invariance 
Eq. (\ref{henkelpred}) could be connected with the 
Fourier transformability of $R_{\bf x} (t,s)$, concluding that some insight
could be obtained by looking at the full $q$ dependence of $R_q(t,s)$. 
This dependence was too cumbersome to be carried out in the two-loop 
computation of Model A and it is still a difficult task for the 
Model C dynamics at one-loop.
In a forthcoming work we analyze the full $q$ dependence of $R_q(t,s)$
for a $\varphi^3$ theory, showing that no problem arises with the Fourier 
transform. 
The nature of such a disagreement should be probably found in the limits of
applicability of local scale invariance. 

\section*{Acknowledgments}
We would thank Malte Henkel for useful comments, suggestions, and 
correspondences.



\begin{references}
\bibitem{foot}
We are not interested here in the problem of phase ordering dynamics (see 
Ref.~\cite{bray} for a review), so we will not consider the case of a quench
to $T<T_c$, that instead is considered in \cite{coars}
\bibitem{foot1.2}
We remark that the exponent $a$ introduced here and in 
Refs.~\cite{cg-02a,cg-02b,cg-02c} is not the same as that used in 
Refs.~\cite{gl-000,gl-02b,hpgl-01}
\end{references}
\end{document}